%% file: main.tex
\documentclass[conference]{IEEEtran}
\IEEEoverridecommandlockouts

\usepackage{cite}
\usepackage{amsmath,amssymb,amsfonts}
\usepackage{algorithmic}
\usepackage{graphicx}
\usepackage{overpic}
\usepackage[hypcap=false]{subcaption}
\usepackage{textcomp}
\usepackage{xcolor}
\usepackage{hyperref}
\hypersetup{
    colorlinks=true,
    citecolor=blue,   
    linkcolor=cyan,    
    filecolor=green,      
    urlcolor=cyan
}
\usepackage{booktabs, multicol, multirow}
\usepackage{pgffor}
\definecolor{lavender}{RGB}{156, 107, 250}

\def\BibTeX{{\rm B\kern-.05em{\sc i\kern-.025em b}\kern-.08em
    T\kern-.1667em\lower.7ex\hbox{E}\kern-.125emX}}

\begin{document}

\title{Denoising Textures is a Different Ballgame Altogether!}



\author{
\IEEEauthorblockN{
\begin{minipage}{\textwidth}\centering
Saptarshi Mandal$^\star$,
Oindrila Haldar$^\star$,
Chirayata Bhattacharya$^\star$,
Chandra Sekhar Seelamantula
\end{minipage}
}
\IEEEauthorblockA{
\begin{minipage}{\textwidth}\centering
\textit{Department of Electrical Engineering}, \\
\textit{Indian Institute of Science},
Bangalore, India \\
\texttt{\{saptarshiman, oindrilah, chirayatab\}@alum.iisc.ac.in, css@iisc.ac.in}
\end{minipage}
}

\and

\IEEEauthorblockN{
\begin{minipage}{0.9\columnwidth}\centering
Zhiwei Xu,
Thrasyvoulos N. Pappas
\end{minipage}
}
\IEEEauthorblockA{
\begin{minipage}{0.9\columnwidth}\centering
\textit{Department of Electrical and Computer Engineering} \\
Northwestern University,
Evanston, Illinois, USA \\
\texttt{zhiweiXu2050@u.northwestern.edu\\ t-pappas@northwestern.edu}
\vspace{-1em}
\end{minipage}
}

\and

\IEEEauthorblockN{
\begin{minipage}{0.9\columnwidth}\centering
Gaurav Sharma
\end{minipage}
}
\IEEEauthorblockA{
\begin{minipage}{0.9\columnwidth}\centering
\textit{Department of Electrical and Computer Engineering} \\
University of Rochester,
Rochester, NY, USA \\
\texttt{gaurav.sharma@rochester.edu}
\vspace{-1em}
\end{minipage}
}
}

\maketitle
\let\thefootnote\relax\footnotetext{$^\star$Equal contribution.}

\input{sections/abstract}

\input{sections/intro}

\input{sections/related_works}

\input{sections/evaluation_methodology}

\input{sections/results}

\input{sections/conclusion}

\bibliographystyle{IEEEbib}
\bibliography{strings, refs}

\end{document}

%% file: sections/abstract.tex
\begin{abstract}
The human brain effortlessly deciphers colors and textures from visual stimuli and reconstructs the world around us, yet preserving such details remains challenging in image denoising. While significant progress has been made in generic image denoising, the problem of texture denoising has remained relatively under-explored. Fine details, such as intricate patterns and repetitive structures, often suffer from over-smoothing even in deep learning based approaches for denoising. In this paper, we analyze the behavior of state-of-the-art (SOTA) denoisers on texture datasets and show that performance improvements are possible only by fine-tuning the denoisers on the texture datasets. Surprisingly, fine-tuning on texture datasets result in the models catastrophically forgetting the original dataset. These findings highlight the need for specialized treatment of textures in denoising. Since denoising quality directly impacts downstream tasks such as generative modelling, we argue that dedicated texture-aware denoisers are essential for faithful texture reconstruction and synthesis.
\end{abstract}

%% file: sections/intro.tex
\section{Introduction}
\label{sec:intro}

Most imaging techniques inevitably introduce noise, which is further exacerbated by environmental conditions, varying lighting conditions and transmission errors. This results in distortion, loss of image information, and overall degradation in visual quality. The presence of noise adversely impacts subsequent image/video processing/analysis tasks. Consequently, image denoising is a critical component of modern image processing pipelines. Image denoising is central to denoising diffusion models~\cite{DDPM, Residual_DDPM} which are SOTA generative models.

The goal of image denoising is to recover the pristine image ${\bf x} \in \mathbb{R}^m$ from its noisy observation ${\bf y} \in \mathbb{R}^m$. However, since perfect restoration is not possible, the aim is to estimate $\bf x$ as accurately as possible. Noise introduces distortions that obscure important details, making estimation of $\bf x$ a challenging task. The most commonly used observation model is $\bf y = x + n $, where ${\bf n} \in \mathbb{R}^m$ denotes additive white Gaussian noise. The objective is to develop a denoising method that produces an estimate of the clean image $\bf \hat{x}$ that is as close as possible to the ground truth $\bf x$.

\begin{figure}[!t]
    \begin{minipage}[b]{0.46\linewidth}
      \centering
      \includegraphics[width=\textwidth, trim=0 80 0 0, clip]{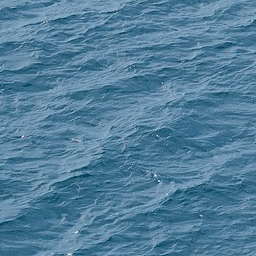}
    \end{minipage}
    \hspace{1.5pt}
    \begin{minipage}[b]{0.46\linewidth}
        \centering
        \includegraphics[width=\textwidth, trim=0 80 0 0, clip]{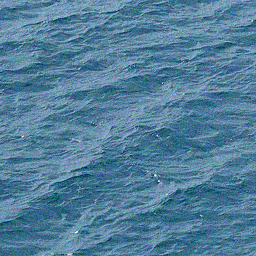}
    \end{minipage}

    \begin{minipage}[b]{0.46\linewidth}
      \centering
      \vspace{1.5pt}
      \includegraphics[width=\textwidth, trim=50 50 100 175, clip]{Images/Introduction/gt_0.05.png} 
      \centerline{Reference} \par
      \centerline{}
    \end{minipage}
    \hspace{1.5pt}
    \begin{minipage}[b]{0.46\linewidth}
      \centering
      \vspace{1.5pt}
      \includegraphics[width=\textwidth, trim=50 50 100 175, clip]{Images/Introduction/noisy_0.05.png}
      \centerline{Noisy} \par
      \centerline{\textbf{26.03 dB}, \textbf{0.801}}
    \end{minipage}

    \begin{minipage}[b]{0.46\linewidth}
      \centering
      \includegraphics[width=\textwidth, trim=0 80 0 0, clip]{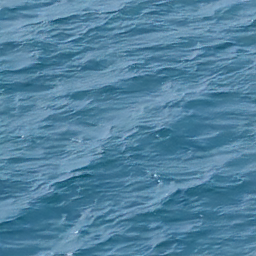}
    \end{minipage}
    \hspace{1.5pt}
    \begin{minipage}[b]{0.46\linewidth}
      \centering
      \includegraphics[width=\textwidth, trim=0 80 0 0, clip]{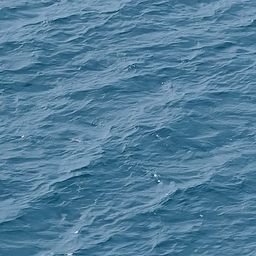}
    \end{minipage}

    \begin{minipage}[b]{0.46\linewidth}
      \centering
      \vspace{1.5pt}
      \includegraphics[width=\textwidth, trim=50 50 100 175, clip]{Images/Introduction/MIRNet_0.05.png}
      \centerline{MIRNet \cite{MIRNet}} \par
      \centerline{\textbf{27.69 dB}, \textbf{0.819}}
    \end{minipage}
    \hspace{1.5pt}
    \begin{minipage}[b]{0.46\linewidth}
      \centering
      \vspace{1.5pt}
      \includegraphics[width=\textwidth, trim=50 50 100 175, clip]{Images/Introduction/MIRNet_finetuned_0.05.png}
      \centerline{MIRNet$^\dagger$} \par
      \centerline{\textbf{32.93 dB}, \textbf{0.946}}
    \end{minipage}
    \caption{\textit{Comparison of a noisy and denoised texture using pretrained MIRNet \cite{MIRNet} and its finetuned version (MIRNet$^\dagger$). PSNR and SSIM values improve after finetuning.}}
    \captionsetup{skip=17pt}
    \label{fig:intro_comparison}
    \vspace{-10pt}
\end{figure}

\begin{figure*}[t]
    \centering
    \captionsetup[subfigure]{labelformat=empty} 

    \subcaptionbox{\shortstack{Noisy \\ 26.06 dB}}[0.14\textwidth]{
        \includegraphics[width=\linewidth]{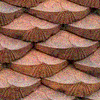}
    }\hfill
    \subcaptionbox{\shortstack{CycleISP \\ 26.31 dB}}[0.14\textwidth]{
        \includegraphics[width=\linewidth]{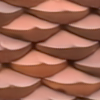}
    }\hfill
    \subcaptionbox{\shortstack{MPRNet \\ 27.22 dB}}[0.14\textwidth]{
        \includegraphics[width=\linewidth]{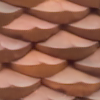}
    }\hfill
    \subcaptionbox{\shortstack{CAT \\ 28.96 dB}}[0.14\textwidth]{
        \includegraphics[width=\linewidth]{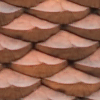}
    }\hfill
    \subcaptionbox{\shortstack{Restormer \\ 29.18 dB}}[0.14\textwidth]{
        \includegraphics[width=\linewidth]{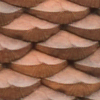}
    }\hfill
    \subcaptionbox{\shortstack{MIRNet \\ 27.53 dB}}[0.14\textwidth]{
        \includegraphics[width=\linewidth]{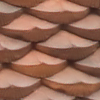}
    }\hfill
    \subcaptionbox{\shortstack{NAFNet \\ 26.19 dB}}[0.14\textwidth]{
        \includegraphics[width=\linewidth]{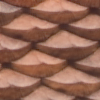}
    }\hfill
    \subcaptionbox{\shortstack{Reference \\ $\infty$ dB}}[0.14\textwidth]{
        \includegraphics[width=\linewidth]{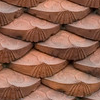}
    }\hfill
    \subcaptionbox{\shortstack{CycleISP$^\dagger$ \\ 33.97 dB}}[0.14\textwidth]{
        \includegraphics[width=\linewidth]{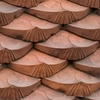}
    }\hfill
    \subcaptionbox{\shortstack{MPRNet$^\dagger$ \\ 34.01 dB}}[0.14\textwidth]{
        \includegraphics[width=\linewidth]{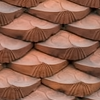}
    }\hfill
    \subcaptionbox{\shortstack{CAT$^\dagger$ \\ 33.80 dB}}[0.14\textwidth]{
        \includegraphics[width=\linewidth]{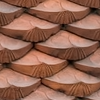}
    }\hfill
    \subcaptionbox{\shortstack{Restormer$^\dagger$ \\ 33.67 dB}}[0.14\textwidth]{
        \includegraphics[width=\linewidth]{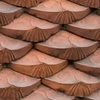}
    }\hfill
    \subcaptionbox{\shortstack{MIRNet$^\dagger$ \\ 33.63 dB}}[0.14\textwidth]{
        \includegraphics[width=\linewidth]{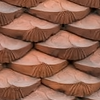}
    }\hfill
    \subcaptionbox{\shortstack{NAFNet$^\dagger$ \\ 33.43 dB}}[0.14\textwidth]{
        \includegraphics[width=\linewidth]{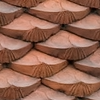}
    }
    \caption{\textit{Visual comparison of various denoising models on the \textbf{NUUR-Texture500}~\cite{nuur} dataset.} \textbf{Top row}: noisy and pre-trained model results, \textbf{bottom row}: reference and (Model$^\dagger$) fine-tuned models. PSNR is shown for each result relative to the reference.}
    \label{fig:denoising_comparison}
    \captionsetup{skip=10pt}
    \vspace{-10pt}
\end{figure*}

Despite significant progress, most deep learning techniques focus on generic image denoising, with limited analysis on specialized classes such as textures. The objective of this paper is to fill this gap. \textit{In image processing, texture can be defined as a function of spatial variation of the brightness intensity of the pixels}~\cite{texture_defn}. Natural textures involve fine details and repetitive elements with some degree of stochasticity that are particularly sensitive to noise. 

As a result, most conventional denoising methods tend to blur intricate patterns in textures, making texture preservation more challenging than generic image denoising. Figure \ref{fig:intro_comparison} illustrates that a deep learning model trained on standard image denoising datasets performs poorly on texture images, both quantitatively and qualitatively. Fine-tuning the model on a texture dataset enhances the performance, highlighting the need for texture-aware denoising strategies.

In this paper, our objective is to carry out a systematic study of the generalizability of deep learning techniques for denoising texture images. As argued in ~\cite{STSIM2}, SSIM is inadequate for objective assessment of the quality of texture. Therefore, we employ alternative measures such as STSIM1~\cite{STSIM2}, STSIM2~\cite{STSIM2}, MS-SSIM~\cite{ms_ssim}, CW-SSIM~\cite{cw_ssim} and DISTS~\cite{dists} in addition to PSNR and SSIM. Among these metrics, STSIM1 and STSIM2 are specifically designed for quality assessment of texture patterns.

%% file: sections/related_works.tex
\section{Related Works}
\label{sec:related_works}

Classical image denoising methods include Gaussian smoothing, bilateral filtering, non-local means~\cite{nlm}, wavelet thresholding, and BM3D~\cite{dabov2007image}. However, these approaches have been surpassed by SOTA deep learning techniques. Jain and Seung~\cite{CNN} first demonstrated the effectiveness of Convolutional Neural Networks (CNNs) for denoising, showing that CNNs could match or even surpass the performance of the then traditional models. Subsequently, with the introduction of SRCNN~\cite{SRCNN} and ARCNN~\cite{ARCNN} for super-resolution and JPEG artifact reduction, respectively, CNN-based methods have achieved impressive performance and have since become mainstream in the field of image restoration.

Denoising Autoencoders (DAEs)~\cite{denoising_autoencoders} introduced learning noise distributions to recover clean images via mapping inputs to meaningful intermediate representations. This was followed by CNN-based models such as DnCNN~\cite{denoising_cnn}, which leveraged deep residual learning~\cite{residual_connections} to effectively remove Gaussian noise. Originally proposed to address degradation of accuracy with the depth of the network, it was also found to improve performance in image classification and object detection.

CNN-based denoising methods process images either at full resolution, preserving spatial details but lacking global contextual, or at reduced resolutions, capturing semantics but losing spatial accuracy. To address this trade-off, \textit{Multi-Scale Image Restoration Network (MIRNet)}~\cite{MIRNet} employs parallel multi-scale feature extraction with attention-based aggregation, while \textit{Multi-Stage Progressive Image Restoration Network (MPRNet)}~\cite{MPRNet} integrates contextual features extracted from an encoder-decoder network with a high-resolution branch that preserves local details. This ensures both spatial precision and strong contextual representation.

\begin{figure*}[t]
    \centering
    \captionsetup[subfigure]{labelformat=empty} 

    \subcaptionbox{\shortstack{Noisy \\ 25.19 dB}}[0.14\textwidth]{
        \includegraphics[width=\linewidth]{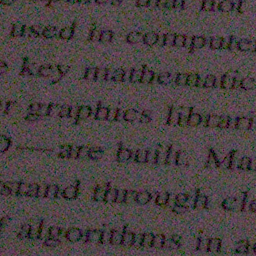}
    }\hfill
    \subcaptionbox{\shortstack{CycleISP \\ 41.00 dB}}[0.14\textwidth]{
        \includegraphics[width=\linewidth]{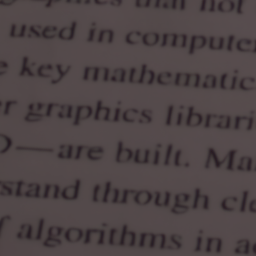}
    }\hfill
    \subcaptionbox{\shortstack{MPRNet \\ 41.27 dB}}[0.14\textwidth]{
        \includegraphics[width=\linewidth]{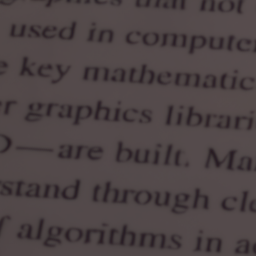}
    }\hfill
    \subcaptionbox{\shortstack{CAT \\ 42.51 dB}}[0.14\textwidth]{
        \includegraphics[width=\linewidth]{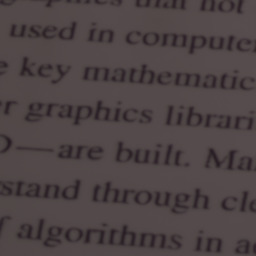}
    }\hfill
    \subcaptionbox{\shortstack{Restormer \\ 42.75 dB}}[0.14\textwidth]{
        \includegraphics[width=\linewidth]{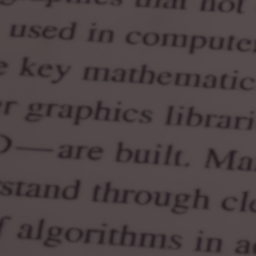}
    }\hfill
    \subcaptionbox{\shortstack{MIRNet \\ 41.54 dB}}[0.14\textwidth]{
        \includegraphics[width=\linewidth]{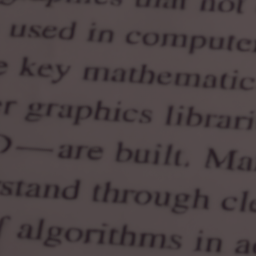}
    }\hfill
    \subcaptionbox{\shortstack{NAFNet \\ 45.59 dB}}[0.14\textwidth]{
        \includegraphics[width=\linewidth]{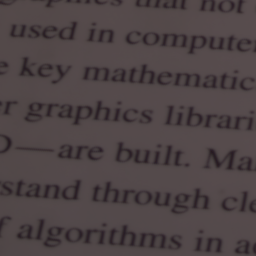}
    }\hfill
    \subcaptionbox{\shortstack{Reference \\ $\infty$ dB}}[0.14\textwidth]{
        \includegraphics[width=\linewidth]{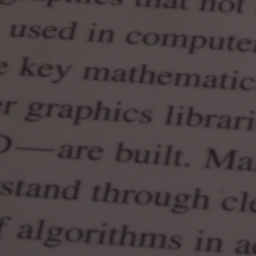}
    }\hfill
    \subcaptionbox{\shortstack{CycleISP$^\dagger$ \\ 29.93 dB}}[0.14\textwidth]{
        \includegraphics[width=\linewidth]{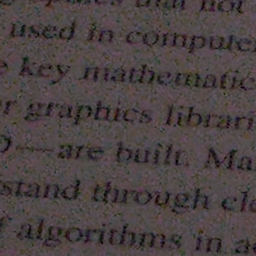}
    }\hfill
    \subcaptionbox{\shortstack{MPRNet$^\dagger$ \\ 28.54 dB}}[0.14\textwidth]{
        \includegraphics[width=\linewidth]{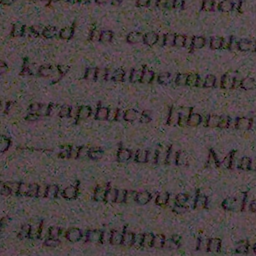}
    }\hfill
    \subcaptionbox{\shortstack{CAT$^\dagger$ \\ 28.29 dB}}[0.14\textwidth]{
        \includegraphics[width=\linewidth]{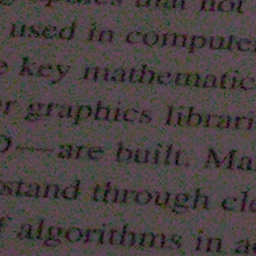}
    }\hfill
    \subcaptionbox{\shortstack{Restormer$^\dagger$ \\ 29.97 dB}}[0.14\textwidth]{
        \includegraphics[width=\linewidth]{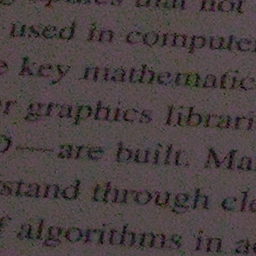}
    }\hfill
    \subcaptionbox{\shortstack{MIRNet$^\dagger$ \\ 29.17 dB}}[0.14\textwidth]{
        \includegraphics[width=\linewidth]{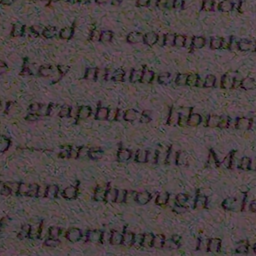}
    }\hfill
    \subcaptionbox{\shortstack{NAFNet$^\dagger$ \\ 33.17 dB}}[0.14\textwidth]{
        \includegraphics[width=\linewidth]{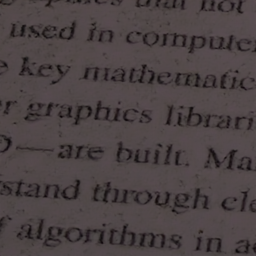}
    }
    \caption{\textit{Visual comparison of denoising models on the \textbf{SIDD}~\cite{SIDD}.} \textbf{Top row}: noisy and pre-trained models, \textbf{bottom row}: reference and (Model$^\dagger$) fine-tuned models. PSNR (dB) shown w.r.t. the reference.}
    \label{fig:denoising_comparison_2}
    \captionsetup{skip=10pt}
\end{figure*}

Most image-denoising algorithms are developed and evaluated on synthetic data, typically generated under the assumption of additive white Gaussian noise (AWGN). However, AWGN does not accurately represent real-world camera noise. \textit{CycleISP}~\cite{CycleISP} addresses this by modeling the camera imaging pipeline in both forward and reverse directions, to generate realistic image pairs in both RAW and sRGB spaces.

Improving performance often increases computational cost. \textit{Nonlinear Activation Free Network (NAFNet)}~\cite{NAFNet} mitigates this by removing or replacing nonlinear activations with multiplications, achieving improved performance with reduced complexity.

To overcome CNN limitations such as restricted receptive fields and static filters, transformer-based models have been proposed. \textit{Restoration Transformer (Restormer)}~\cite{Restormer} leverages global context for SOTA restoration performance. But transformers with global attention are computationally expensive. To reduce the high computational cost of global attention, \textit{Cross Aggregation Transformer (CAT)}~\cite{CAT} introduced rectangle-window self-attention. This approach applies horizontal and vertical rectangle window attention in parallel across different heads, expanding the attention area and enabling feature aggregation across windows.

%% file: sections/evaluation_methodology.tex
\section{Experimental Setup}

Despite the strong performance of modern CNN and transformer-based denoisers on synthetic Gaussian noise and real-world datasets such as SIDD~\cite{SIDD}, their performance when the distribution of images shift to texture-rich ones remains unaddressed. It is important to understand the dissimilarities between natural images and textures. While textures are components of natural images, they differ structurally when considered independently. As defined by Portilla \& Simoncelli~\cite{portilla2000}, ``\textit{An image of visual texture is spatially homogeneous and typically contains repeated structures, often with some random variation (e.g., random positions, orientations or colors)."} In contrast, natural images have been observed to predominantly consist of smooth (low-pass) regions~\cite{huang1999statistics}. This characteristic is fundamentally different from textures, which may or may not exhibit smoothness. The presence of high-frequency spectral information, which is also present in noise, is what makes texture denoising challenging. In this paper, the objective is to validate and benchmark the performance of SOTA deep denoisers, specifically on textures.

To benchmark the performance, SOTA denoising models, which have demonstrated exceptional performance in real-world image denoising, were selected --- MIRNet~\cite{MIRNet}, MPRNet~\cite{MPRNet}, CAT~\cite{CAT}, CycleISP~\cite{CycleISP}, NAFNet~\cite{NAFNet}, and Restormer~\cite{Restormer}. These models were evaluated for denoising performance on a texture dataset using the same noise levels as the original SOTA models. Subsequently, the models are fine-tuned on the texture dataset, and their performance is evaluated again. Fine-tuning was adopted instead of training from scratch, as training these models from scratch requires a substantially larger dataset, which was not available for the present study.

\vspace{-5pt}
\subsection{Datasets}
\label{subsec:dataset}

For qualitative and quantitative evaluation of the pre-trained and fine-tuned models, patches were extracted from the {\bf NUUR-Texture500} dataset~\cite{nuur} which contains 500 diverse and spatially consistent and uniform texture images spanning from regular patterns to stochastic ones. From each image in the {\bf NUUR-Texture500} dataset~\cite{nuur}, ten $256\times256$ patches were randomly sampled, yielding 5000 patches in total, split into 75\% for training and 25\% for validation. To simulate real-world noise conditions, noisy patches were generated by adding white Gaussian noise, with the noise standard deviation ($\sigma$) determined by estimating the average variance of noise in the noisy images of the {\bf SIDD} dataset~\cite{SIDD}. Our goal was to match the noise characteristics of the synthetic noisy texture patches with those found in real-world datasets such as SIDD, enabling a fair evaluation of denoising performance. 

\subsection{Implementation Details}
\label{subsec:implementation}
Six Nvidia A6000 GPUs were used to fine-tune the models using PyTorch. The fine-tuning of the denoisers is done using Adam Optimizer ($\beta_1=0.9$, $\beta_2=0.999$) for 100 epochs with a batch size of 16. The initial learning rate is set to $1\times10^{-4}$ and a cosine annealing strategy is used to steadily reduce the learning rate to $5\times10^{-5}$ during training. The loss function is chosen in accordance with the network architecture and the same objective used during training is retained for fine-tuning (e.g. $\mathcal{L}_1$ loss for CAT and \textit{Charbonnier} loss for MPRNet).

\subsection{Evaluation Metrics}
\label{subsec:eval}

\begin{table*}[t!]
\centering
\scriptsize
\begin{tabular}{l|c|c|c|c|c|c|c|c}
\toprule
\textbf{Model} & \textbf{PSNR}$\uparrow$ & \textbf{SSIM}$\uparrow$ & \textbf{MS-SSIM}$\uparrow$ & \textbf{CW-SSIM}$\uparrow$ & \textbf{DISTS}$\downarrow$ & \textbf{STSIM1}$\uparrow$ & \textbf{STSIM2}$\uparrow$ & \textbf{LPIPS}$\downarrow$ \\
\midrule
Noisy & 26.23 $\pm$ 0.36 & 0.860 $\pm$ 0.123 & 0.976 $\pm$ 0.024 & 0.962 $\pm$ 0.077 & 0.157 $\pm$ 0.076 & 0.989 $\pm$ 0.023 & 0.941 $\pm$ 0.052 & 0.1286 $\pm$ 0.1488 \\
\midrule
CAT~\cite{CAT} & 27.73 $\pm$ 4.542 & 0.834 $\pm$ 0.149 & 0.956 $\pm$ 0.074 & 0.931 $\pm$ 0.127 & 0.165 $\pm$ 0.077 & 0.954 $\pm$ 0.092 & 0.926 $\pm$ 0.062 & 0.1904 $\pm$ 0.1778 \\
CAT$^\dagger$ & 34.17 $\pm$ 3.389 & 0.959 $\pm$ 0.032 & 0.990 $\pm$ 0.009 & 0.973 $\pm$ 0.054 & 0.062 $\pm$ 0.034 & 0.997 $\pm$ 0.006 & 0.949 $\pm$ 0.050 & 0.0304 $\pm$ 0.0283 \\
\midrule
CycleISP~\cite{CycleISP} & 25.52 $\pm$ 3.759 & 0.781 $\pm$ 0.192 & 0.933 $\pm$ 0.099 & 0.872 $\pm$ 0.212 & 0.239 $\pm$ 0.089 & 0.929 $\pm$ 0.136 & 0.911 $\pm$ 0.077 & 0.3047 $\pm$ 0.2425 \\
CycleISP$^\dagger$ & 34.20 $\pm$ 3.346 & 0.959 $\pm$ 0.032 & 0.998 $\pm$ 0.005 & 0.973 $\pm$ 0.053 & 0.062 $\pm$ 0.034 & 0.998 $\pm$ 0.005 & 0.948 $\pm$ 0.050 & 0.0306 $\pm$ 0.0296 \\
\midrule
MIRNet~\cite{MIRNet} & 25.23 $\pm$ 5.765 & 0.721 $\pm$ 0.204 & 0.917 $\pm$ 0.104 & 0.913 $\pm$ 0.144 & 0.232 $\pm$ 0.081 & 0.909 $\pm$ 0.127 & 0.909 $\pm$ 0.070 & 0.3425 $\pm$ 0.2464 \\
MIRNet$^\dagger$ & 33.88 $\pm$ 3.171 & 0.958 $\pm$ 0.033 & 0.990 $\pm$ 0.009 & 0.972 $\pm$ 0.054 & 0.064 $\pm$ 0.034 & 0.997 $\pm$ 0.005 & 0.950 $\pm$ 0.049 & 0.0319 $\pm$ 0.0309 \\
\midrule
MPRNet~\cite{MPRNet} & 25.24 $\pm$ 5.674 & 0.721 $\pm$ 0.207 & 0.915 $\pm$ 0.108 & 0.919 $\pm$ 0.134 & 0.236 $\pm$ 0.080 & 0.911 $\pm$ 0.131 & 0.909 $\pm$ 0.068 & 0.3432 $\pm$ 0.2590 \\
MPRNet$^\dagger$ & 34.17 $\pm$ 3.297 & 0.959 $\pm$ 0.032 & 0.990 $\pm$ 0.009 & 0.972 $\pm$ 0.054 & 0.062 $\pm$ 0.033 & 0.997 $\pm$ 0.005 & 0.949 $\pm$ 0.050 & 0.0303 $\pm$ 0.0291 \\
\midrule
NAFNet~\cite{NAFNet} & 23.93 $\pm$ 6.399 & 0.715 $\pm$ 0.235 & 0.877 $\pm$ 0.187 & 0.859 $\pm$ 0.205 & 0.251 $\pm$ 0.095 & 0.931 $\pm$ 0.086 & 0.909 $\pm$ 0.068 & 0.3397 $\pm$ 0.2412 \\
NAFNet$^\dagger$ & 33.59 $\pm$ 3.429 & 0.953 $\pm$ 0.064 & 0.987 $\pm$ 0.048 & 0.967 $\pm$ 0.082 & 0.064 $\pm$ 0.044 & 0.994 $\pm$ 0.055 & 0.946 $\pm$ 0.068 & 0.0362 $\pm$ 0.0728 \\
\midrule
Restormer~\cite{Restormer} & 26.81 $\pm$ 5.213 & 0.817 $\pm$ 0.148 & 0.949 $\pm$ 0.073 & 0.936 $\pm$ 0.118 & 0.186 $\pm$ 0.070 & 0.957 $\pm$ 0.066 & 0.923 $\pm$ 0.059 & 0.2327 $\pm$ 0.1808 \\
Restormer$^\dagger$ & 33.94 $\pm$ 3.144 & 0.958 $\pm$ 0.032 & 0.990 $\pm$ 0.009 & 0.972 $\pm$ 0.054 & 0.062 $\pm$ 0.032 & 0.997 $\pm$ 0.004 & 0.950 $\pm$ 0.051 & 0.0305 $\pm$ 0.0287 \\
\bottomrule
\end{tabular}
\caption{\textit{Denoising performance (mean $\pm$ std) on \textbf{NUUR-Texture500 dataset}\cite{nuur} using PSNR, SSIM, MS-SSIM, CW-SSIM, DISTS, STSIM1, STSIM2 and LPIPS}. Dagger ($^\dagger$) indicates fine-tuned models.}
\vspace{-8pt}
\label{table:Comparison_other_metrics}
\end{table*}

Performance is evaluated using both fidelity and perceptual metrics such as PSNR, SSIM, MS-SSIM~\cite{ms_ssim}, CW-SSIM~\cite{cw_ssim}, STSIM1~\cite{STSIM2}, STSIM2~\cite{STSIM2}, DISTS~\cite{dists} and LPIPS~\cite{zhang2018unreasonable}.

\noindent{\bf Structural Texture Similarity (STSIM-1)} captures texture similarity using local image statistics by replacing the structure term in SSIM with directional first-order autocorrelation of neighboring subband coefficients computed horizontally and vertically denoted by $\rho_{\mathbf{x}}(0,1)$ and $\rho_{\mathbf{x}}(1,0)$ respectively. 
\begin{equation*}
    c_{\mathbf{x}, \mathbf{y}}(0,1) = 1 - 0.5 \left| \rho_{\mathbf{x}}(0,1) - \rho_{\mathbf{y}}(0,1) \right|
\end{equation*}  
\begin{equation*}
    c_{\mathbf{x}, \mathbf{y}}(1,0) = 1 - 0.5 \left| \rho_{\mathbf{x}}(1,0) - \rho_{\mathbf{y}}(1,0) \right|
\end{equation*}
The metric is calculated for each subband or patch and consolidated as follows:
\[{\text{STSIM-1}}(\mathbf{x}, \mathbf{y}) = (l(x,y) \cdot c(x,y) \cdot  c_{\mathbf{x}, \mathbf{y}}(0,1) \cdot c_{\mathbf{x}, \mathbf{y}}(1,0))^\frac{1}{4}.\] 

\noindent{\bf STSIM-2} extends STSIM-1 by incorporating cross-band correlations. The crossband correlation between subbands $m$ and $n$ is computed as follows: 
\[ \rho_{|\mathbf{x}|}^{m,n} (0,0) =
    \frac{\mathbb{E} \left\{ \left[ |x^m(i,j)| - \mu_{|\mathbf{x}|}^m \right] 
    \left[ |x^n(i,j)| - \mu_{|\mathbf{x}|}^n \right] \right\}}
    {\sigma_{|\mathbf{x}|}^m \sigma_{|\mathbf{x}|}^n}.\]
Combining the STSIM-1 values over all subbands gives
\[{\text{STSIM-2}}(\mathbf{x}, \mathbf{y}) =\frac{
\sum_{m=1}^{N_b} q_{\text{STSIM-1}}^m (\mathbf{x}, \mathbf{y}) +
\sum_{i=1}^{N_c} \mathbf{c}_{\mathbf{x}, \mathbf{y}}^{m_i, n_i} (0,0)
}{N_b + N_c},\]
where $N_b$ denotes the number of subbands, and $N_c$ accounts for the cross-correlations considering various orientations.\\

\noindent{\bf STSIM-M} Different image subbands statistically vary in terms of subband statistics like mean, variance, horizontal and vertical auto-correlation and cross-correlation across subbands. A feature vector \(F_{\mathbf{x}}\) (and \(F_{\mathbf{y}}\)) is formed by stacking all these features for images \(\mathbf{x}\) and \(\mathbf{y}\), respectively. STSIM-M is then defined as the Mahalanobis distance~\cite{mahalanobis} between these features, considering that they are mutually uncorrelated and combined.
\[
\text{STSIM-M}(F_{\mathbf{x}}, F_{\mathbf{y}}) = \sqrt{(F_{\mathbf{x}} - F_{\mathbf{y}})^H M (F_{\mathbf{x}} - F_{\mathbf{y}})}
\]

\noindent{\bf Deep Image Structure and Texture Similarity (DISTS)})~\cite{dists} is a deep learning-based metric that assesses the similarity of structural and texture information in images. It uses high-level perceptual features extracted from different layers of a pre-trained VGG network. 

%% file: sections/results.tex
\section{Results}
\label{sec:results}

\begin{table}[t]
\centering
\begin{tabular}{l|c|c}
\toprule
\textbf{Model} & \textbf{PSNR}$\uparrow$ & \textbf{SSIM}$\uparrow$ \\
\midrule
Noisy & 27.86 $\pm$ 5.17 & 0.52 $\pm$ 0.23 \\
\midrule
MIRNet~\cite{MIRNet} & 42.59 $\pm$ 4.08 & 0.96 $\pm$ 0.06 \\
MIRNet$^\dagger$ & 32.68 $\pm$ 6.75 & 0.72 $\pm$ 0.26 \\
\midrule
MPRNet~\cite{MPRNet} & 42.60 $\pm$ 4.10 & 0.96 $\pm$ 0.06 \\
MPRNet$^\dagger$ & 32.48 $\pm$ 6.80 & 0.72 $\pm$ 0.26 \\
\midrule
CycleISP~\cite{CycleISP} & 42.21 $\pm$ 4.12 & 0.95 $\pm$ 0.06 \\
CycleISP$^\dagger$ & 33.98 $\pm$ 7.33 & 0.76 $\pm$ 0.27 \\
\midrule
CAT~\cite{CAT} & 43.28 $\pm$ 4.36 & 0.96 $\pm$ 0.06 \\
CAT$^\dagger$ & 32.68 $\pm$ 7.48 & 0.73 $\pm$ 0.27 \\
\midrule
NAFNet~\cite{NAFNet} & 44.83 $\pm$ 4.55 & 0.96 $\pm$ 0.05 \\
NAFNet$^\dagger$ & 37.34 $\pm$ 5.55 & 0.88 $\pm$ 0.16 \\
\midrule
Restormer~\cite{Restormer} & 43.36 $\pm$ 4.39 & 0.96 $\pm$ 0.06 \\
Restormer$^\dagger$ & 32.80 $\pm$ 6.42 & 0.72 $\pm$ 0.25 \\
\bottomrule
\end{tabular}
\caption{\textit{Original vs. fine-tuned ($^\dagger$) model performance on \textbf{SIDD}~\cite{SIDD}.}}
\label{table3}
\captionsetup{skip=10pt}
\vspace{-10pt}
\end{table}

Figure~\ref{fig:denoising_comparison} presents a qualitative comparison of performance of various denoising models on texture images taken from the NUUR-Texture500 dataset. The corresponding quantitative results, averaged over the validation set, are reported in Table~\ref{table:Comparison_other_metrics} in terms of PSNR, SSIM, STSIM1, STSIM2, and other metrics. The fine-tuned models demonstrate significant improvements across various performance metrics. Although the models were fine-tuned on pointwise losses (e.g. $\mathcal{L}_1$), they also exhibited improvements in non-pointwise perceptual metrics such as MS-SSIM, CW-SSIM, STSIM1, STSIM2 and DISTS.\\
\indent Fine-tuned models show a considerable increase in PSNR, ranging from \textbf{+6.44} to \textbf{+9.66 dB} and subjective metric such as LPIPS shows an improvement of around \textbf{-0.16} to \textbf{-0.31} over the baseline models. This result clearly indicates that the baseline models are not adequately equipped to handle textures, and it is only upon fine-tuning, that the performance of the models can be improved significantly. Fine-tuned models also performed better across all the other structural similarity metrics (SSIM variants) and DISTS, indicating the importance of fine-tuning for achieving structurally and perceptually superior denoising performance. The performance improvement happens across multiple scales as indicated by the MS-SSIM metric. The STSIM1 and STSIM2 metrics, which assess structural and multi-scale structural similarity specifically for textures, also show consistent improvements (\textbf{+0.065} and \textbf{+0.034} on average, respectively).\\
\indent Next, we carried out an evaluation of the fine-tuned denoising models on the original natural image datasets. The performance is reported in Table~\ref{table3}. We observe that the models fine-tuned on the texture dataset do not perform as well on the original dataset. A visual demonstration is presented in Figure~\ref{fig:denoising_comparison_2}. This result shows that, in the process of adaptation to the texture dataset, the models have \textit{catastrophically forgotten}~\cite{catastrophic_forgetting} the original dataset. The incongruence in performance suggests that natural images and texture images have low-dimensional embeddings that overlap minimally with each other. Thus, one can infer that textures have to be handled separately compared to other natural images. A unified denoising model that handles both texture images and other natural images equally well is an exciting direction for future research. 

%% file: sections/conclusion.tex
\section{Conclusion}
\label{sec:Conclusion}
In this study, we benchmarked the performance of SOTA image-denoisers on a texture dataset. We have shown that \textit{off-the-shelf denoisers achieve subpar denoising performance on texture images} compared to generic natural images. In our benchmarking, we have also compared the denoising performance using several metrics like STSIM1, STSIM-2 that capture textural nuances better than SSIM. \textit{We have fine-tuned the models on the texture dataset and showed that the texture denoising performance improves, but at the expense of significant degradation in performance on the original dataset}. The study highlights the need for better denoisers that can handle textures better, either separately or along with general image datasets. However, it is to be noted that since the models were trained on point-by-point (e.g. $\mathcal{L}_1$) losses, further efforts are required to allow more extensive use of perceptual metrics, such as STSIM-M and STSIM-VGG, that do not operate on a point-by-point basis.

The next step in the study would be to design unified denoisers and train them on perceptual losses that are not point-by-point which can allow for a more holistic texture analysis. We believe that such unified denoisers can play a crucial role in the generation of natural images with rich texture in a diffusion or flow-based modelling framework.